# GÉANT Software Maturity Model

Zarko Stanisavljevic, Bartosz Walter, Maja Vukasovic,
Andrijana Todosijevic, Maciej Labedzki, Marcin Wolski

*Abstract* — GÉANT project is an example of a large organization with around 30 software projects and around 20 software development teams. Software development teams consist of many skilled associates coming from all members National Research and Education Networks. Three main issues that are common for all these software development teams and their members are: geographical distribution, scattered manpower percentage, and parallel involvement in other high priority projects in their native organizations. This paper presents a novel software maturity model that is designed specifically for GÉANT software development teams and aims to address the described issues.

*Keywords* — CMMI, GÉANT, Software Development Methodologies, Software Maturity Model

## I. Introduction

EVERY software product and every service has to be developed with the highest attainable quality in order to fulfill the needs of its users. Moreover, software products that reach the production phase must be developed and delivered within the planned budget and schedule. This is especially hard to achieve within an organization that is relying on a geographically distributed software development teams (SWDTs), such as GÉANT [1]. Environment, organization and management of work within the distributed SWDTs can be a complex process, as developers may have varying work methodologies in terms of style, experience, expectations and approach. Despite recent advances in the theory and practice of software engineering, planning and evaluating the implementation of software products within a distributed environment, with only partial and changing involvement of team members, remain exceedingly challenging. The inherent complexity of such projects commonly leads to suboptimal software quality, unnecessary technical debt and expensive software evolution processes.

In an organization such as GÉANT, product portfolio is relying on a large number of software projects developed with the effort of many skilled associates coming from all members National Research and Education Networks (NRENs) [2]. The NREN community is involved in collaborative software development activities, focusing on delivering software products that provide advanced services, including but not limited to connectivity, network management, trust and identity, cloud solutions, etc. within the GÉANT project.

Elaborate analysis about the GÉANT software development practices showed that there is a need expressed by the GÉANT software development community to address the issue of optimization of the software development process [3]. This issue could be addressed by providing the SWDTs with guidance on adopting and using software development methodologies in an efficient and effective way, along with creating a reference method for assessing the teams' performance.

This paper presents a novel maturity model framework that can be used to guide software development teams in GÉANT toward their desired goals. Unlike software methodologies, maturity models propose a framework that defines goals to be achieved. A SWDT can choose the specific methods of achieving the goals. Additionally, maturity models allow for monitoring the progress made by the SWDTs, by measuring their performance in key areas. They also provide the SWDTs with a method for objective evaluation of its performance, indicating directions for improvement.

The rest of the paper is organized as follows. Section II gives more details about the motivation for designing the new maturity model. Section III presents an overview of existing maturity models and gives an explanation why none of the existing models could be used for GÉANT SWDTs. Section IV introduces the new maturity model and describes next steps that are going to be taken in order to evaluate the effectiveness of the proposed model.

## II. Motivation

One of the objectives of GÉANT project [4] is ensuring that production services offered by GÉANT are of acceptable quality and managed efficiently, with relevant procedures and processes in place and all relevant documentation available. The GÉANT product portfolio contains software projects with different maturity, size and target domains: starting from the prototype (proof of concept) solutions, through the pilot applications that

This work was financed by © GÉANT Association on behalf of the GN4-2 project. The research leading to these results has received funding from the European Union's Horizon 2020 research and innovation programme under Grant Agreement No. 731122 (GN4-2).

Zarko Stanisavljevic and Maja Vukasovic are with the School of Electrical Engineering, University of Belgrade, Bul. kralja Aleksandra 73, 11120 Belgrade, Serbia (phone: +381-11-3218392; e-mail: zarko.stanisavljevic@etf.bg.ac.rs, maja.vukasovic@etf.bg.ac.rs).

Corresponding Bartosz Walter is with the Poznań Supercomputing and Networking Center, ul. Jana Pawła II 10, 61-139 Poznan, Poland (phone: +48-618585143; e-mail: bartek.walter@man.poznan.pl).

Andrijana Todosijevic is with the Serbian Academic Network - AMRES, Bul. kralja Aleksandra 90, 11000 Belgrade, Serbia (phone: +381-11-7158942; e-mail: andrijana.todosijevic@amres.ac.rs).

Maciej Labedzki and Marcin Wolski are with the Poznan Supercomputing and Networking Center, ul. Jana Pawła II 10, 61-139 Poznan, Poland (phone: +48-61-8585143; e-mail: labedzki@man.poznan.pl, marcin.wolski@man.poznan.pl).

usually target a closed group of users, to completed products, supporting the delivery of services in production. Their users mostly include GÉANT partners and their member institutions, researchers, students and educators, who need high quality, reliable services and infrastructure to support their work or studies.

The SWDTs in the GÉANT project constitute a federated community, consisting of experienced and skilled software engineers coming from different NRENs and GÉANT. Most of GÉANT's developers are included in several projects and dedicate only a limited amount of their time to each task they are entrusted with [3]. When a developer distributes small amounts of his/her time across many different software projects, it negatively affects team's productivity and performance. Lower quality code produces lower value product and ongoing costs when it comes to software maintenance. At the same time inefficient, non-intuitive and unreliable software not only lacks value, it may also generate costs for its end users [5].

Setting up an effective software development process can optimize cost and value of a product. Even small teams, which are the majority in the GÉANT [3], are aware of the benefits coming from the agile software development methodologies (SDMs) [6]. Although the GÉANT software community is well prepared for that type of work process and the teams have proved their ability to finish their projects successfully, there are still some areas that need improvements. For example, the level of adoption of SDMs varies among the teams. They have struggled with obstacles, such as time pressure or lack of experience in the consistent adoption of best practices. Some reasons for problems come from the geographical distribution of the team members, scattered manpower percentage, and parallel involvement in other high priority projects. As a result, efforts towards optimizing the development process by adopting standards and best practices are not sufficiently coordinated, which leads to adopting solutions that are not optimal and not well-suited to the specifics of GÉANT.

In this paper we aim to address the described issues by providing the SWDTs with an appropriate software maturity model that would serve two purposes: (1) it could be used for assessing the teams' performance, and (2) to guide the team in adopting best practices concerning the development process in an efficient and effective way.

## III. RELATED WORK

The growing importance of processes in software industry has been recognized quite early. Nolan [7] proposed a staged, six-level approach to implementing IT in business. Works by several other authors, e.g., Yourdon [8] and DeMarco [9] also indicated a need for establishing a common ground for software development processes and practices, based on experience and empirical research.

Finally, a process maturity framework, proposed by Humphrey in 1988 [10], inspired researchers from Software Engineering Institute at CMU, who defined Capability Maturity Model (CMM) [11], a five-stage scheme that identifies major areas of responsibilities and capabilities in software development. CMM, initially oriented toward military applications, quickly evolved into an industry de facto standard and gave rise to several spin-off models, tailored to domains other than software, e.g., People CMM [12]. The success of CMM resulted in a further refinement (called Capability Maturity Model Integration, CMMI) and adaptations to other domains (e.g., CMMI for Development, Acquisition or Services).

Since its inception, CMMI has become a viable and widely accepted method of evaluating the capabilities of software vendors: the list of organizations that have undergone the CMMI appraisal [13] includes thousands of entries, representing both world-wide leaders, but also small or medium-sized software houses from most countries.

Critics of CMM/CMMI focus on its strict process-orientation, ignoring the human factor [14]. A major risk is that organizations may choose to technically fulfill the goals of the model, rather than properly address and optimize existing processes. On the other hand, by following a CMMI road map, an organization does not become more mature *per se*, and vice versa: CMMI is not the only way towards improving maturity.

Although CMMI is considered a formal approach, the notion of maturity has been also transferred to and implemented in other, more informal environments. Open-source communities developed a number of models that reflect the capability of the projects, teams and individuals, e.g., Capgemini-Open Source Maturity Model (C-OSMM) [15], Qualification and Selection of Open Source [16] (QSOS) or OMM for open-source communities [17]. They address issues not covered by CMMI, like communication, or re-formulate existing ones, e.g., the approach to requirements definition and management.

Also agile methodologies have adopted the concept of maturity. Agile Maturity Model (AMM) [18] focused on defining levels of agility, which increasingly address the common practices and values starting from basic ones, e.g., planning and requirements management, up to managing uncertainty and defect prevention.

Comparative studies on software development methods indicated that their convergence is not only possible, but also beneficial. The same applies to maturity evaluation: Paulk found eXtreme Programming [19] largely compliant with CMM Level 2, and Schweigert at al. analyzed agile models from a SPICE perspective [20]. Also a SLR on the using CMMI together with agile approaches [21] shows that agility helps in getting to levels 2 and 3 of CMMI, although it should be supplemented with more formal practices as well. The collected evidence clearly indicates the need for institutionalization of practices that have been found both essential and beneficial for software development performance [22].

## IV. THE PROPOSED MATURITY MODEL

As described in the previous part in literature and in use there are a lot of maturity models, customized to software development models' needs. CMMI focuses on the technical processes and represents a more formal approach, while models for open-source embrace also soft skills and effective communication within teams. Neither of these approaches fully fits the needs of GÉANT as a software organization. On one hand CMMI is too heavyweight and generic, while other models are too specific to agile software development. Additionally, no model addresses the specifics of GÉANT: geographical distribution of teams and scattered involvement of individuals in the teams. The Maturity Model for GÉANT SWDTs needs to seek for a balance between these two described approaches. The core of the model could be based on CMMI, but the result model should also incorporate several goals and activities specific to agile development methods, which would produce a model tailored to the needs of an innovation-driven, distributed organization.

Maturity models have either descriptive or prescriptive purpose, i.e., the intention to describe an as-is-situation or to define a road map for improvements. In our model we chose prescriptive approach, based on the needs gathered from GÉANT SWDTs. We also chose to use the continuous representation over staged one for the GÉANT maturity model. Measuring the software development team's maturity in different areas without giving a single, overall score of maturity, thus avoiding unnecessary benchmarking and rather focusing on the improvement is enabled in that way. This approach also provides more flexible representation, since it allows for changes to be made easier in the future.

*A. Components*

There are five main components of the proposed MM, as depicted in Figure 1:

- Key Process Area (KPA, PA) – a cluster of related practices representing a single process or a family of processes, which, when implemented collectively, satisfy a set of goals considered important for embracing the area and pursuing improvements in it,
- Specific Goal (SG) – an objective related to a single process, identified as an essential element of the KPA,
- Parameter – a weighted measurable attribute associated with an SG,
- GÉANT PLM (Product Lifecycle Management) position – a link between a given KPA and a stage in the GÉANT PLM,
- Scoring System – a mechanism for evaluating the parameters in a quantifiable manner.

At the moment, the proposed MM for GÉANT comprises of five KPAs: Requirements Engineering, Design and Implementation, Quality Assurance, Team Organization, and Software Maintenance. These are selected to cover the entire GÉANT development process according to GÉANT PLM. Therefore, each KPA is associated with a corresponding PLM position.

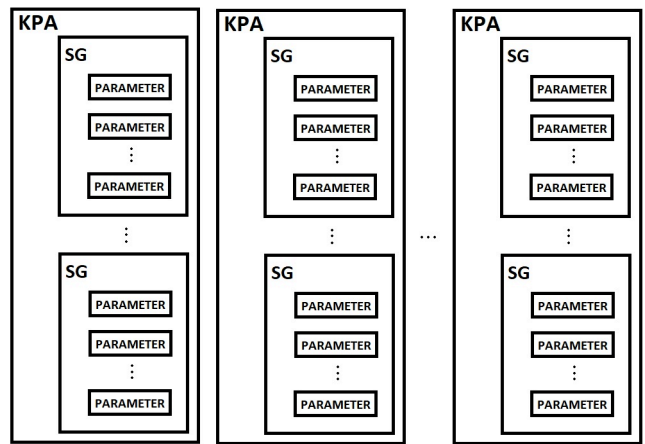

Figure 1. Design of the proposed MM

Each KPA defines a number of SGs. A unique pool of measurable parameters has been defined in order to measure the extent of fulfilling the SG. Specific parameters are associated with the appropriate SGs in a manner that each SG can have one or more parameters from the unique pool and that each parameter from the unique pool can be assigned to one or more SGs. The parameters are assigned to one of five categories: Process Quality, Estimation Accuracy, Product Quality, Team Quality, and Technology Quality.

Finally, to evaluate the parameters within SGs, a scoring system is devised. Each parameter could be evaluated at one of three levels:

- 0 – not available / not used / not defined / unaware;
- 1 – implicit (known and used, but not written down);
- 2 – explicit (written down, institutionalized);

*B. Methods*

Once the scores are collected, they need to be aggregated and interpreted. Currently, we consider two different methods for score aggregation and mapping the results onto maturity evaluation. The methods differ with respect to their complexity: the compensatory method is simpler, but also has side effects due to possible compensation between various SGs, while the two-tier method is more sophisticated, but discriminates between basic and advanced goals. Below, we provide short descriptions of the methods.

*1) Compensatory method*

In this method, each parameter is evaluated according to the scoring system. Then, scores get aggregated at the PA's level by simple adding. Weights at the SG's level could represent the relative importance of each parameter within the SG, and should be calibrated before the maturity model is implemented. Finally, the SG is adjusted to fit a value between 0 and 100, which represents the percentage of completeness for that SG. The score at the PA level can be calculated as a mean value for all SG values that belong to that PA.

The proposed interpretation of the maturity evaluation is as follows:

- value 0-30 – initial level
- value 30-60 – intermediate level
- values 60-80 – advanced level
- values 80-100 – optimizing level

*2) Two-tier method*

Again, each parameter is evaluated according to the scoring system. Next, scores get aggregated at the PA's level in two separate categories: basic SGs and advanced SGs. Then, each PA is described by two scores, representing the basic and advanced SGs. The score at the PA level is adjusted to fit a value between 0 and 100, which describes the percentage of completeness for that PA.

The proposed interpretation of the maturity evaluation is as follows:

- basic 0-50 – initial level
- basic 50-80 and advanced 0-20 – intermediate level
- basic 80-100 and advanced 20-50 – advanced level
- basic 80-100 and advanced 50-100 – optimized level

Both methods allow for weighting the relative importance of specific SGs, but in a different manner. The proposed methods of evaluation will enable teams to observe their strong and weak points and improve their organization in the important areas as estimated.

*C. Plan of evaluation*

Before implementing the maturity model, it needs to be evaluated and approved. Implementation of a maturity model into practice faces various difficulties. First, the distributed structure of GÉANT teams promotes their independence and self-organization. That hinders comparison of results, but also makes the teams reluctant to be evaluated. For that reason, we proposed a plan of evaluation, intended to overcome these issues and get valuable feedback from SWDTs. The plan involves following steps:

- **Instruction** – Selected representatives of SWDT are instructed about the scope and meaning of the maturity model. The representatives make a diverse sample of different team profiles.
- **Review by selected SWDT representatives** – The representatives perform guided review of the maturity model. A co-author of the model is present to respond to their questions. Feedback is collected and summarized.
- **Revision of the model** – Authors revise the model, using the feedback collected in the previous step.
- **Pilot evaluation on selected SWDT** – The revised model is evaluated on the selected team(s). Results are analyzed, and proposals for change are formulated.
- **Second revision of the model** – Authors revise the model again, incorporating the results of the pilot evaluation.
- **Dissemination of the model within GÉANT** – The model definition is disseminated among SWDTs.
- **Implementation** – The model is applied in practice to provide the teams with reports on their maturity.

ACKNOWLEDGMENT

The authors would like to thank all current and former members of the SA2 T1 team, all participants of the 2017 side meeting at the GÉANT symposium, all participants of the 2018 TF SMD proposal meeting and all GÉANT participants who had a chance to look at the idea presented in the paper and give us their valuable feedback.